\documentclass[12pt]{article}
\usepackage{graphicx}
\usepackage{url}
\usepackage{amsfonts}
\usepackage[T1]{fontenc}
\usepackage{float}
\usepackage[utf8]{inputenc}
\usepackage{subfigure}
\usepackage{epsfig,epsf}
\usepackage{amsbsy}
\usepackage{textcomp}
\usepackage{mathrsfs}
\usepackage{amsmath,amssymb}
\usepackage{latexsym}
\usepackage{xcolor}
\usepackage{color}

\usepackage{float}

\begin{document}

\title{Closer look at  some sum  uncertainty relations\\ and related relations: \\ Their weak points}
\author{
Krzysztof  Urbanowski\\
Institute of Physics,
University of Zielona G\'{o}ra,  \\
ul. Prof. Z. Szafrana 4a,
65-516 Zielona G\'{o}ra,
Poland\\
\hfill\\
e--mail:  K.Urbanowski@if.uz.zgora.pl}

\maketitle

\begin{abstract}
We analyze the weak and critical points of some state--dependent uncertainty relations that follow from the inequalities for the norms of vectors in the Hilbert space of states of a quantum system. Uncertainty relations for sums of standard deviations, for sums of variances, and other relations between standard deviations or variances are studied. The obtained results are compared with the conclusions obtained in similar cases using the standard Heisenberg--Robertson uncertainty relation.
We also show that, in critical cases, the right--hand sides of sum and stronger sum uncertainty relations provide similar information regarding lower bounds on the left--hand sides of these relations as the Heisenberg--Robertson uncertainty relations do.
New reverse uncertainty relations for a sum of variances are derived.
\end{abstract}

\noindent
{\bf Key words:} Uncertainty relations, Sum uncertainty relations, Reverse uncertainty relations, Inequalities for norm of vectors.

\section{Introduction }
In classical physics describing physical phenomena in our macroscopic world, physical quantities are described by functions in which the variables are positions and velocities (or momentum). The accuracy with which we measure these physical quantities depends only on the perfection of our measuring instruments and is independent of whether we simultaneously measure one or more physical quantities. In turn, in quantum mechanics describing processes occurring in the microscopic world, physical quantities are represented by observables, i.e. self--adjoint operators acting in the state space ${\cal H}$ of a given physical system.
And here, unlike the macroscopic world, if physical quantities under studies
are represented by non--commutative operators (say $A$ and $B$), $AB \neq BA$,
then unlike the classical world, it is not possible to unequivocally state what values the observables $A$ and $B$ will take when we try to simultaneously measure their values. This effect explains
the famous uncertainty principle discovered by Heisenberg. This principle is the one of characteristic and the most important consequences of the quantum mechanics. It
indicates the fundamental differences between classical and quantum world.
Heisenberg's uncertainty principle states that the following relation must hold for the position and momentum \cite{H}:
\begin{equation}
\Delta_{\phi} x\ \cdot \Delta_{\phi} p_{x}\,\geq \,\frac{\hbar}{2}, \label{H1}
\end{equation}
where $\Delta_{\phi} x$ and $\Delta_{\phi} p_{x}$ are the standard (root--mean--square) deviations \cite{M}. The inequality (\ref{H1}) is the most known form of  the uncertainty principle and it was 
improved mathematically by Kennard \cite{Ken} and 
generalized by Robertson \cite{R} for any two  non--commuting observables $A$ and $B$,
\begin{equation}
\Delta_{\phi} A \cdot \Delta_{\phi} B\;\geq\;\frac{1}{2} \left|\langle [A,B] \rangle_{\phi} \right|,\label{R1}
\end{equation}
Here
\begin{equation}
\Delta_{\phi} F \equiv \sqrt{\langle F^{2}\rangle_{\phi} - \langle F\rangle_{\phi}^{2}}. \label{DF}
\end{equation}
(where $F = A$ and $B$) denotes the standard deviations, $|\phi\rangle$ is a  state of the system under considerations, $\langle \psi|\phi\rangle$ is the inner product in the state space ${\cal H}$ of the system, ($|\psi\rangle, |\phi\rangle \in {\cal H}$), $\langle F\rangle_{\phi} = \langle \phi|F\,\phi\rangle \equiv \langle \phi|F|\phi\rangle$ is the average (expected) value of  the observable $F$ in the state $|\phi\rangle$, provided that $|\langle\phi|F|\phi \rangle |< \infty$.
 The observable $F$ is represented by hermitian operator $F$ acting in a Hilbert space ${\cal H}$ of states.
 Here  $[A,B] = AB - BA$  and it is assumed that  states $|\phi\rangle$ under consideration are  normalized: $\langle \phi|\phi\rangle \stackrel{\rm def}{=} \|\,|\phi\rangle \|^{2} =1$. (In Eq. (\ref{H1})  $F$ stands for position  and momentum operators $x$ and $p_{x}$). 
(In  general, the uncertainty relation (\ref{R1}) depends on the state of the system and is called state--dependent, and only this type of uncertainty relations will be considered in this paper).
Various properties of these relations have been analyzed, e.g.,  by Schr\"{o}dinger \cite{S} who proposed the stronger form of (\ref{R1}), and many other authors, (see, e. g. \cite{Cow} --- \cite{Bi3} and so on), and 
 among others, also in the author's papers \cite{ku-min,ku1}.

	The inspiration for this paper comes from some recent publications in which the authors claim that the Heisenberg--Robertson (HR) uncertainty relation for two non--commuting observables,  $A$ and $B$,
has some imperfections
because it does not provide any information about the standard deviations when the system is in one of the eigenstates of one of these observables (see, eg. \cite{Mac,Bin,Xiao,Pat}): 
This is because the right--hand side of inequality (\ref{R1}) then equals zero (as does one of the factors on the left-hand side of that inequality).
It is claimed in these papers that much better uncertainty relations are the new ones of the type of  ,,sum uncertainty relations'', because the above--mentioned ,,weak point'' of the HR relation does not appear  in these new ones. For this reason, in the following parts of the paper we will analyze in detail the properties of  uncertainty relations of types discussed in \cite{Mac,Bin,Xiao,Pat}, 
that is of types
derived using the triangle inequality and related inequalities, as well as their generalizations.
In particular we will examine 
properties 
of such relations in the case
when the quantum system is in one of the 
the eigenstates of the analyzed pair of non--commuting observables.
Also the properties of uncertainty relations of this type in other critical cases, i.e. in cases where the right hand side of the HR uncertainty relation  is equal to zero, will be analyzed in the further parts of this paper.
In this paper, we continue the considerations initiated in \cite{ku-rev1,ku-edu}; the analysis presented there is significantly extended and deepened here.

It should be noted here that other types of sum uncertainty relations are also considered in the literature (see, e.g., \cite{Schwo,Ab,Bush3,Bush4,Bush5} and similar papers), but they will not be analyzed in the present work.

The paper is organized as follows: Sec. 2  contains, for the convenience of readers,  a  derivation of the Heisenberg--Robertson uncertainty relation and analysis of some its properties. In Sec. 3 readers will find selected inequalities for vectors from the state space ${\cal H}$, which are useful in studying the uncertainty principles. The derivation of the  uncertainty relations for the sum of variances and similar related uncertainty relations  is given in Sec. 4.
In this Section, the reader will also find both well--known and new inequalities satisfied by standard deviations (or variances), 
as well as an analysis of the weaknesses of the reverse uncertainty relation derived by  Mondal, Bagchi, and  Pati \cite{Deb}. 
At the end of this Section, the reader will also find a derivation of new reverse uncertainty relations for sums of variances.
Section 5 contains some new results:
we analyze therein the critical  points of the uncertainty relations derived in Sec. 4. Discussion of the results presented in this paper and final remarks are given in Sec. 6.

\section{Heisenberg--Robertson uncertainty relation}
As we have already mentioned, the inequality (\ref{R1}) holds for each pair of non-commuting observables $A$ and $B$ represented by Hermitian operators acting in the state space ${\cal H}$ of the physical system under study,
such that $[A,B]$ exists and $|\phi\rangle \in {\cal D}(AB) \bigcap {\cal D}(BA) \subset {\cal H}$, (${\cal D}({\cal O})$ denotes the domain of an operator $\cal O$ or of a product of operators).
Note that, the relation (\ref{R1}) results from basic assumptions of the quantum theory and from the geometry of Hilbert space. Indeed,
taking into account that
in the general case for an observable $F$ the standard deviation is defined as follows
\begin{equation}
\Delta_{\phi} F = \|\delta_{\phi} F|\phi\rangle\| \geq 0, \label{dF}
\end{equation}
where  $\delta_{\phi} F = F - \langle F\rangle_{\phi}\,\mathbb{I}$,
and defining
\begin{equation}
\delta_{\phi} A = A - \langle A\rangle_{\phi}\,\mathbb{I},\;\;{\rm and,} \;\;
\delta_{\phi} B = B - \langle B\rangle_{\phi}\,\mathbb{I},
\end{equation}
 one finds that
\begin{equation}
[A,B] \equiv [\delta_{\phi} A, \delta_{\phi} B],
\end{equation}
for all $|\phi\rangle \in {\cal D}(AB) \bigcap {\cal D}(BA)$.

Taking all this into account, we get  that
\begin{eqnarray}
\left|\langle \phi| [A,B] |\phi \rangle \right|^{2} &\equiv & \left|\langle \phi| [\delta_{\phi} A,\delta_{\phi} B] |\phi \rangle \right|^{2} \label{[dA,dB]} \\
& =  & 4\,\left|\,{\rm Im.}\left[\langle \phi| \delta_{\phi} A \;\delta_{\phi} B |\phi \rangle \right]\, \right|^{2} \label{Im[]} \\
&\leq& 4 \,\left| \,\langle \phi| \delta_{\phi} A \;\delta_{\phi} B |\phi \rangle \,\right|^{2} \label{C00} \\
& \leq & 4 \,\left\| \delta_{\phi} A |\phi \rangle \right \|^{2}\,\cdot\,\left \|\delta_{\phi} B|\phi \rangle \right\|^{2}  \nonumber \\
& \equiv & 4 \;(\Delta_{\phi}A)^{2}\,\cdot \,(\Delta_{\phi}B)^{2}, \label{H-ku}
\end{eqnarray}
which reproduces inequality (\ref{R1}).  There is $ (\Delta_{\phi} A)^{2} \equiv \left\|\delta_{\phi} A|\phi\rangle \right\|^{2}$ and $ (\Delta_{\phi} B)^{2} \equiv \left\|\delta_{\phi} B|\phi\rangle \right\|^{2}$, (Relations (\ref{dF}) and (\ref{DF}) are equivalent). Here the Schwarz inequality was used:
\begin{equation}
\left| \,\langle \phi| \delta_{\phi} A \;\delta_{\phi} B |\phi \rangle \,\right|^{2} \leq  \,\left\| \delta_{\phi} A |\phi \rangle \right \|^{2}\,\cdot\,\left \|\delta_{\phi} B|\phi \rangle \right\|^{2}.
\label{Schw1}
\end{equation}

	There are physicists who consider the weak side of inequality (\ref{R1}) to be that its right--hand side is equal to zero when it is calculated for a vector $|\phi\rangle$ being  an eigenvector of one of the observables $A$ or $B$, which makes this inequality trivial, (see, e.g. \cite{Mac,Bin,Xiao}).
	Indeed,  if to
 assume that $[A,B]\neq 0$ and $|\phi \rangle = |\psi_{b}\rangle$ is a normalized eigenvector of  $B$ for the eigenvalue $b$, that is that
  \begin{equation}
  B|\psi_{b}\rangle = b |\psi_{b}\rangle. \label{B=b}
  \end{equation}
  and that $|\psi_{b}\rangle \in {\cal D}(AB) \bigcap {\cal D }(BA)$, 
  then one  finds that $\langle \psi_{b}|AB|\psi_{b}\rangle $ \linebreak
    $= b \,\langle \psi_{b}|A|\psi_{b}\rangle$ and thus
  \begin{eqnarray}
  	\langle [A,B] \rangle_{\psi_{b}} &\equiv& \langle \psi_{b}|  [A,B] |\psi_{b} \rangle \nonumber \\
  	&=& \langle \psi_{b}|  AB |\psi_{b} \rangle - \langle \psi_{b}|  BA |\psi_{b} \rangle \equiv 0, \label{[A,B]=0}
  \end{eqnarray}
i.e., for  $|\phi \rangle = |\psi_{b}\rangle$, the right--hand side of (\ref{R1}) vanishes.
The  assumption (\ref{B=b}) immediately implies also that
  \begin{equation}
  \langle \psi_{b}|B|\psi_{b} \rangle = b,\; \;{\rm and,}\;\; \delta_{\psi_{b}} B|\psi_{b}\rangle \equiv 0. \label{B=b+1}
  \end{equation}
   As a result we have that
   \begin{equation}
 \Delta_{\psi_{b}}B = \left\| \delta_{\psi_{b}} B |\psi_{b} \rangle \right \| =0, \label{DBb=0}
 \end{equation}
which, together with (\ref{[A,B]=0}),  means that if $|\phi\rangle$ is an eigenstate of the observable $B$, $|\phi\rangle = |\psi_{b}\rangle$, then  we have no  information from (\ref{R1}) on $\Delta_{\psi_{b}}A$.
Due to the alleged weakness of the HR uncertainty relation, the remainder of this paper will focus primarily on comparing the properties of the relations proposed in \cite{Mac,Bin,Xiao} with those of the HR relation.

\section{Preliminaries: useful inequalities}

As a ,,cure'' for the above described ,,weakness'' of the uncertainty relation (\ref{R1}) a new uncertainty relation for the sum of the variances of two non-commuting observables, $A$ and $B$, has been proposed (see e.g. \cite{Mac,Pat} and also \cite{Hof} and earlier mentioned other papers).  The starting points for the ,,sum uncertainty relations'' are some inequalities taking place in the Hilbert state space.
The basis of various types of "sum uncertainty relations" are: the triangle inequality,
\begin{equation}
\left\|\;| \psi_{A}\rangle \right\| + \left\|\;| \psi_{B}\rangle \right\|\;\geq \; \left\|\;| \psi_{A}\rangle \; + \;| \psi_{B}\rangle \right\|,\label{t1}
\end{equation}
the triangle inequality of the second kind (see, e.g. \cite{Jam}),
\begin{equation}
\left\|\;| \psi_{A}\rangle \right\|^{2} + \left\|\;| \psi_{B}\rangle \right\|^{2}\;\geq \; \frac{1}{2}\left\|\;| \psi_{A}\rangle \; + \;| \psi_{B}\rangle \right\|^{2},\label{t2}
\end{equation}
which follows from the parallelogram law,
\begin{equation}
 2\left(  {\left\| \;|\psi_{A}  \rangle \right\|}^{2} +
{ \left\|\; |\psi_{B}  \rangle \right\|}^{2} \right)  =  { \left\|\; |\psi_{A}  \rangle +  |\psi_{B}  \rangle \right\| }^{2} + { \left\| \;|\psi_{A}  \rangle -  |\psi_{B}  \rangle \right\|}^{2}, \label{par}
\end{equation}
and other related inequalities. Let us note that from  (\ref{par}) there follows another inequality,
\begin{equation}
\left\|\;| \psi_{A}\rangle \right\|^{2} + \left\|\;| \psi_{B}\rangle \right\|^{2}\;\geq \; \frac{1}{2}\left\|\;| \psi_{A}\rangle \; - \;| \psi_{B}\rangle \right\|^{2}.\label{t2a}
\end{equation}
Still further useful inequalities can be obtained using the following identities,
\begin{eqnarray}
\left\|\,|\psi_{A}\rangle\, - \,|\psi_{B}\rangle\right\|^{2} &=&\left\|\,|\psi_{A}\rangle\right\|^{2}\,+\,\left\|\,|\psi_{B}\rangle\right\|^{2}\,
-\,2 \,\Re(\langle \psi_{A}|\psi_{B}\rangle), \label{id1}\\
\left\|\,|\psi_{A}\rangle\, + \,|\psi_{B}\rangle\right\|^{2} &=&\left\|\,|\psi_{A}\rangle\right\|^{2}\,+\,\left\|\,|\psi_{B}\rangle\right\|^{2}\,
+\,2 \,\Re(\langle \psi_{A}|\psi_{B}\rangle). \label{id2}
\end{eqnarray}
Directly from (\ref{id1}) using the Schwarz inequality we find that (for details see \cite{ku-rev1}),
\begin{equation}
2\,\left\|\;| \psi_{A}\rangle \right\|\left\|\;| \psi_{B}\rangle \right\|
 \;\geq \; \left\|\;| \psi_{A}\rangle \right\|^{2}\, + \, \left\|\;| \psi_{B}\rangle \right\|^{2}\,
- \,  \left\| \;|\psi_{A}  \rangle -  |\psi_{B}  \rangle \right\|^{2}. \label{t1^2<}
\end{equation}
In this way, upon rearranging this inequality, we obtain
\begin{equation}
\left\| \;|\psi_{A}  \rangle -  |\psi_{B}  \rangle \right\|^{2}\,+\,
2\,\left\|\;| \psi_{A}\rangle \right\|\left\|\;| \psi_{B}\rangle \right\|
 \;\geq \; \left\|\;| \psi_{A}\rangle \right\|^{2}\, + \, \left\|\;| \psi_{B}\rangle \right\|^{2}.  \label{new-ineq}
\end{equation}

Note that the reverse triangle inequality immediately follows from inequality (\ref{new-ineq}). Indeed, using (\ref{t1^2<}), (\ref{new-ineq})
it is easy to find that
\begin{equation}
\left\| \;|\psi_{A}  \rangle -  |\psi_{B}  \rangle \right\|^{2}\;
 \;\geq \;\left( \left\|\;| \psi_{A}\rangle \right\|\, - \, \left\|\;| \psi_{B}\rangle \right\|\right)^{2},
  \label{rev2}
\end{equation}
which is equivalent to the reverse triangle inequality:
\begin{equation}
\left\| \;|\psi_{A}  \rangle -  |\psi_{B}  \rangle \right\|\;
 \;\geq \;\left| \;{\left\|\;| \psi_{A}\rangle \right\|\, - \, \left\|\;| \psi_{B}\rangle \right\|}\;\right|.
  \label{rev3}
\end{equation}

Next, we can use the following property
\begin{equation}
\left(\left\|\,|\psi_{A}\rangle\right\| - \left\|\,|\psi_{B}\rangle \right\|\right)^{2} = \left\|\,|\psi_{A}\rangle\right\|^{2} + \left\|\,|\psi_{B}\rangle\right\|^{2} \,- \,2\left\|\,|\psi_{A}\rangle\right\|\, \left\|\,|\psi_{B}\rangle\right\|\,\geq\, 0. \label{n1}
\end{equation}
and obtain another inequality valid in the Hilbert space of states,
\begin{equation}
\left\|\,|\psi_{A}\rangle\right\|^{2} + \left\|\,|\psi_{B}\rangle\right\|^{2}\;\geq \;2\left\|\,|\psi_{A}\rangle\right\|\, \left\|\,|\psi_{B}\rangle\right\|.  \label{n2}
\end{equation}

Continuing our considerations, let us note that if we use the identities (\ref{id1}), (\ref{id2}),
and then multiply them by themselves on their respective sides,
we obtain that,
\begin{eqnarray}
\left\|\,|\psi_{A}\rangle - |\psi_{B}\rangle\right\|^{2}\;\left\|\,|\psi_{A}\rangle + |\psi_{B}\rangle\right\|^{2} & = &
  \left(\left\|\,|\psi_{A}\rangle\right\|^{2}+\left\|\,|\psi_{B}\rangle\right\|^{2}\right)^{2} \nonumber  \\
  & &\;\;\;\;\;\;\;\;-4\;\left[\Re(\langle \psi_{A}|\psi_{B}\rangle)\right]^{2}. \label{n10}
\end{eqnarray}
Using this result we get finally (see, e.g., \cite{Bar}),
\begin{equation}
\left\|\,|\psi_{A}\rangle - |\psi_{B}\rangle\right\|\;\left\|\,|\psi_{A}\rangle + |\psi_{B}\rangle\right\|\ \leq
\left\|\,|\psi_{A}\rangle\right\|^{2}+\left\|\,|\psi_{B}\rangle\right\|^{2}. \label{Bar1}
\end{equation}

On the other hand, equation (\ref{n10}) can be rewritten as
\begin{eqnarray}
\left\|\,|\psi_{A}\rangle - |\psi_{B}\rangle\right\|^{2} \cdot \left\|\,|\psi_{A}\rangle + |\psi_{B}\rangle\right\|^{2} & + & 4\;\left| \Re(\langle \psi_{A}|\psi_{B}\rangle) \right|^{2} \nonumber \\
 & = &  \left( \left\|\,|\psi_{A}\rangle\right\|^{2} + 
  \left\|\,|\psi_{B}\rangle\right\|^{2}\right)^{2}. \label{n11}
\end{eqnarray}
Hence the conclusion that,
\begin{eqnarray}
\left( \left\|\,|\psi_{A}\rangle\right\|^{2} +
\left\| \,|\psi_{B}\rangle\right\|^{2}\right)^{2} &\leq &
\left\|\,|\psi_{A}\rangle - |\psi_{B}\rangle\right\|^{2} \cdot \left\|\,|\psi_{A}\rangle + |\psi_{B}\rangle\right\|^{2} \nonumber \\
&& + 4\;\left| \langle \psi_{A}|\psi_{B}\rangle \right|^{2} \nonumber \\
   \label{n11a}
\end{eqnarray}
Then, by applying the Schwarz inequality in (\ref{n11a}), we obtain another inequality
\begin{eqnarray}
 \left( \left\|\,|\psi_{A}\rangle\right\|^{2} + 
  \left\|\,|\psi_{B}\rangle\right\|^{2}\right)^{2}\,&\leq&\,\left\|\,|\psi_{A}\rangle - |\psi_{B}\rangle\right\|^{2} \cdot \left\|\,|\psi_{A}\rangle + |\psi_{B}\rangle\right\|^{2} \nonumber \\
  && + 4\,\left\|\; \psi_{A}\rangle\right\|^{2}\cdot \left\|\;|\psi_{B}\rangle \right\|^{2}, \label{n12}
\end{eqnarray} 
or, equivalently.
\begin{eqnarray}
  \left\|\,|\psi_{A}\rangle\right\|^{4} + 
  \left\|\,|\psi_{B}\rangle\right\|^{4}\,&\leq&\,\left\|\,|\psi_{A}\rangle - |\psi_{B}\rangle\right\|^{2} \cdot \left\|\,|\psi_{A}\rangle + |\psi_{B}\rangle\right\|^{2} \nonumber \\
  && + 2\,\left\|\; \psi_{A}\rangle\right\|^{2}\cdot \left\|\;|\psi_{B}\rangle \right\|^{2}, \label{n12a}
\end{eqnarray} 
 The last three inequalities seem not to have been considered in the literature.

It is also worth mentioning here the generalization of the triangle inequality to $N$ vectors,
(see, e. g.,  \cite{Jam,Cer,Hsu}),
\begin{equation}
\sum_{j=1}^{N}\left\|\,|\psi_{j}\rangle \right\|\;\geq\; \left\|\sum_{j=1}^{N}|\psi_{j}\rangle \right\|, \label{t-n}
\end{equation}
which can be used to derive the ,,sum uncertainty relation'' for $N$ non--commuting observables.

At the end of the considerations in this Section, let us recall of the Dunkl--Williams inequality, which is used in the literature to derive the so--called ,,reverse uncertainty relation''.
The Dunkl--Williams
inequality for vectors $|\psi_{A}\rangle, |\psi_{B}\rangle$ from a real or complex inner product space has the following form \cite{Dun,Cer},
\begin{align}
\left\|\, |\psi_{A}\rangle - |\psi_{B}\rangle \right\| & \geq \frac{1}{2} \left( \left\|\,|\psi_{A}\rangle\right\| + \left\|\,|\psi_{B}\rangle\right\|\right) \times  \nonumber \\
 &\;\;\;\;\;\times \left\| \frac{|\psi_{A}\rangle}{\left\| \,|\psi_{A}\rangle\right\|} -  \frac{|\psi_{B}\rangle}{\left\| \,|\psi_{B}\rangle\right\|} \right\|, \label{Dun}
\end{align}
where the condition that  $|\psi_{A}\rangle, |\psi_{B}\rangle$ are nonzero vectors must be satisfied \cite{Dun,Cer}. This inequality was used in \cite{Deb} to find the ,,reverse uncertainty relation''.

\section{Sum and other uncertainty relations and inequalities}

Let us now derive the sum uncertainty relation and analyze some of its properties.
Inserting into (\ref{t1})
\begin{equation}
|\psi_{A}\rangle = \delta_{\phi} A|\phi\rangle,\;\; |\psi_{B}\rangle  = \delta_{\phi} B|\phi\rangle, \label{dA}
\end{equation}
 one gets
 \begin{equation}
 \left\|\;\delta_{\phi} A| \phi\rangle \right\| + \left\|\;\delta_{\phi} B| \phi\rangle \right\|\;\geq \; \left\|\delta_{\phi} A| \phi\rangle \; + \;\delta_{\phi} B| \phi\rangle \right\|. \label{P1}
 \end{equation}
 Next, taking into account  (\ref{dF}), we see that the inequality (\ref{P1}) can be written in terms of standard deviations as
\begin{equation}
\Delta_{\phi}A + \Delta_{\phi}B \geq \Delta_{\phi}(A + B). \label{Pat4}
\end{equation}
It is because simply,
\begin{equation}
\delta_{\phi} A| \phi\rangle \; \pm \;\delta_{\phi} B| \phi\rangle \equiv \delta_{\phi}( A \pm B)| \phi\rangle. \label{d(A+B)}
\end{equation}
The inequality (\ref{Pat4}) is the {\em sum uncertainty relation} derived in \cite{Pat} and discussed in the literature.

The generalization of the  triangle inequality
to $N$ vectors
allows generalizing the sum uncertainty relation (\ref{Pat4}) for two noncommuting observables to the case of $N$ non--commuting observables $\left\{ A_{j} \right\}_{j=1}^{N}$ with $ N \geq 2$ for standard deviations, $ (\Delta_{\phi}A_{1}), (\Delta_{\phi}A_{2}), \ldots , (\Delta_{\phi}A_{N})$,
\begin{equation}
\sum_{j=1}^{N}(\Delta_{\phi}A_{j}) \geq   \,\Delta_{\phi}(\sum_{j=1}^{N} A_{j}). \label{t-n1}
\end{equation}
This type of sum uncertainty relation and its various generalizations  have recently been intensively studied in many papers (see, e. g. \cite{Bin,Xiao,Pat,Son} and many others).


In addition to the sum uncertainty relations for standard deviations, the ,,stronger sum uncertainty relations'' for variances have also been studied recently.
Such uncertainty relations can be obtained from  the triangle inequality of the second kind (\ref{t2}). Indeed,
starting from formula (\ref{t2}) and using (\ref{dA}) we get
\begin{equation}
\left\|\;\delta_{\phi} A| \phi\rangle \right\|^{2} + \left\|\;\delta_{\phi} B| \phi\rangle \right\|^{2}\;\geq \; \frac{1}{2}\left\|\delta_{\phi} A| \phi\rangle \; + \;\delta_{\phi} B| \phi\rangle \right\|^{2}.\label{m1}
\end{equation}
Next, taking into account   (\ref{d(A+B)}),
we can replace inequality (\ref{m1}) by the following one,
\begin{equation}
\left\|\;\delta_{\phi} A| \phi\rangle \right\|^{2} + \left\|\;\delta_{\phi} B| \phi\rangle \right\|^{2}\;\geq \; \frac{1}{2}\left\|\;|\delta_{\phi}( A + B)| \phi\rangle \right\|^{2}.\label{m2}
\end{equation}
Keeping in mind (\ref{dF})  one concludes that the inequality (\ref{m2}) means that
\begin{equation}
(\Delta_{\phi}A)^{2} + (\Delta_{\phi}B)^{2} \geq	 \frac{1}{2} \left(\Delta_{\phi}(A+B)\right)^{2}, \label{M12}
\end{equation}
and this inequality is called ,,the stronger sum uncertainty relations'' and  is
the one of sum of variances uncertainty relations derived in  \cite{Mac}.

Note that there is also a related relation resulting from (\ref{t2a}). Repeating all the steps leading to the inequality (\ref{M12}) we obtain,
\begin{equation}
(\Delta_{\phi}A)^{2} + (\Delta_{\phi}B)^{2} \geq	 \frac{1}{2} \left(\Delta_{\phi}(A-B)\right)^{2}. \label{M12aa}
\end{equation}
This relation was studied, eg. in \cite{strong-2}. In the same way, using (\ref{new-ineq}), we find that
\begin{equation}
(\Delta_{\phi}A)^{2} + (\Delta_{\phi}B)^{2}  \leq \left( \Delta_{\phi}(A-B) \right)^{2} + 2 \Delta_{\phi} A\; \Delta_{\phi} B, \label{rev1a}
\end{equation}
which looks like some form of the reverse uncertainty relation derived and discussed in \cite{Deb,Xiao2,Xiao3} and in other papers. The name ,,reverse uncertainty relation'' appeared for the first time  in \cite{Deb}
and it is based on  the Dunkl--Williams inequality:
Replacing   $|\psi_{A}\rangle, |\psi_{B}\rangle$ in (\ref{Dun})  by (\ref{dA}) and using the definition (\ref{DF}) after some algebra Mondal and co--authors of  \cite{Deb} obtained that
\begin{equation}
(\Delta_{\phi} A)^{2}\,+\,(\Delta_{\phi} B)^{2} \;\leq \;2 \frac{[\Delta_{\phi} (A - B)]^{2} }{\left[1- \frac{ cov_{\phi}(A,B) }{\Delta_{\phi} A\cdot \Delta_{\phi} B}\right]}\,-\,2\,\Delta_{\phi} A\cdot \Delta_{\phi} B, \label{Deb}
\end{equation}
where $ cov_{\phi}(A,B) = {\cal C}_{\phi}^{clas}(A,B)  \stackrel{\rm def}{=} \Re\,[{\mathcal C}_{\phi}(A,B)] $, and
\begin{equation}
{\cal C}_{\phi}(A,B) \stackrel{\rm def}{=} \langle \phi|\delta_{\phi}A\;\delta_{\phi}B|\phi\rangle \equiv \langle AB \rangle_{\phi}\, -\, \langle A\rangle_{\phi}\,\langle B\rangle_{\phi}, \label{C1}
\end{equation}
is a quantum version of the correlation function (see, e. g.,  \cite{Poz,Khr,Mac2} and references in \cite{ku-min}).
(The ${\cal C}_{\phi}^{clas}(A,B)$ is the classical part of $C_{\phi}(A,B)$ and corresponds to the classical correlation function).

 The inequality (\ref{Deb}) is the {\em reverse uncertainty relation} derived in \cite{Deb}. As we can see, the left sides of the inequalities (\ref{rev1a}) and (\ref{Deb}) are the same, the differences occur on their right sides and in this sense we can say that the inequality (\ref{rev1a}) is a possible form of the reverse uncertainty relation.


Using (\ref{Deb})
it is worth emphasizing here that in some cases it may happen that there exists such a state $|\phi\rangle$ that  $B|\phi\rangle = \lambda_{\phi}A|\phi\rangle \neq 0$, where $\lambda_{\phi} = \lambda_{\phi}^{\ast} > 0$. Then  $\Delta_{\phi}B = \lambda_{\phi}\Delta_{\phi}A \neq 0$ and ${\cal C}_{\phi}(A,B) =
{\cal C}_{\phi}(A,\lambda_{\phi}A) \equiv \lambda_{\phi}[\Delta_{\phi}A]^{2} $ which implies that
the inequality (\ref{Deb}) does not provide us with any useful information about the upper bound of its left side by the values taken by the right side of this inequality in this state. Simply, if $\lambda_{\phi} \neq 1$ and  $\Delta_{\phi}B = \lambda_{\phi}\Delta_{\phi}A >  0$ and ${\cal C}_{\phi}(A,B) =  {\cal C}_{\phi}(A,\lambda_{\phi}A) \equiv \lambda{\phi}[\Delta_{\phi}A]^{2}$ then
the denominator in the first term of the right side of this inequality  is equal to zero, and thus this term has an infinite value. If $\lambda_{\phi} = 1$ then the first term of the right side of this inequality has form zero divided by zero.

The method used to derive the above uncertainty relations and applied to other inequalities discussed in Sec. 3 allows us to derive other, similar relations. Thus, from the reverse triangle inequality (\ref{rev3}) we obtain,
\begin{equation}
\Delta_{\phi}(A - B)
 \;\geq \;\left|\Delta_{\phi}A\, - \, \Delta_{\phi}B\right|.  \label{n4}
\end{equation}
Note that the inequalities (\ref{rev1a}) and (\ref{n4}) are  equivalent. So, the relation (\ref{rev1a}) is a different face of relation (\ref{n4}).
To see this, simply square both sides of (\ref{n4}) and the result is the relation (\ref{rev1a}). This observation allows us to find an interesting property of the  relation (\ref{rev1a}). Namely, if the  mean standard deviations (or variances) of observables $A$ and $B$ in the state $|\phi\rangle$ are equal, $\Delta_{\phi}A = \Delta_{\phi}B$, then the left--hand side (\ref{n4}) equals zero, and the inequality (\ref{n4}) takes the trivial form
\begin{equation}
\Delta_{\phi}(A - B)
 \;\geq 0.  \label{n4a}
\end{equation}
One obtains a similar result from the inequality (\ref{rev1a}) by substituting $\Delta_{\phi}A = \Delta_{\phi}B$.

A few more
inequalities:
One can conclude from (\ref{n2}) that (see also \cite{Mac,strong-2}),
\begin{equation}
(\Delta_{\phi}A)^{2} + (\Delta_{\phi}B)^{2} \;\geq\;2\,\Delta_{\phi}A\;\Delta_{\phi}B, \label{n5}
\end{equation}
whereas
from (\ref{Bar1}) that
\begin{equation}
\Delta_{\phi}(A-B)\;\Delta_{\phi}(A+B) \leq \left(\Delta_{\phi}A\right)^{2} +  \left(\Delta_{\phi}B\right)^{2}. \label{Bar2}
\end{equation}
This last sum uncertainty relation was derived in \cite{Bar}.

Inserting $|\psi_{A}\rangle$  and $|\psi_{B}\rangle$ in the form (\ref{dA}) into the the inequality (\ref{n11a}) and
and using (\ref{d(A+B)}), (\ref{C1}) we obtain
\begin{eqnarray}
 \left( \left(\Delta_{\phi}A\right)^{2} +  \left(\Delta_{\phi}B\right)^{2} \right)^{2}\,&\leq&\,
 \left[\Delta_{\phi}(A-B)\;\cdot\;\Delta_{\phi}(A+B)\right]^{2} \nonumber \\
 && + \;\,
 4\;\left| {\cal C}_{\phi}(A,B)\right|^{2}. \label{n13}
\end{eqnarray}
The weaker form of (\ref{n13}) is
\begin{eqnarray}
 \left( \left(\Delta_{\phi}A\right)^{2} +  \left(\Delta_{\phi}B\right)^{2} \right)^{2}\,&\leq&\,
 \left[\Delta_{\phi}(A-B)\;\cdot\;\Delta_{\phi}(A+B)\right]^{2} \nonumber \\
 && + \;\,
 4\;\left(\Delta_{\phi}A\;\cdot\;\Delta_{\phi}B \right)^{2}. \label{n13a}
\end{eqnarray}  
The equivalent form of inequality (\ref{n13a}) can be  obtained from (\ref{n12a}) and looks as follows,
\begin{eqnarray}
 \left(\Delta_{\phi}A\right)^{4} +  \left(\Delta_{\phi}B\right)^{4} \,&\leq&\,
 \left[\Delta_{\phi}(A-B)\;\cdot\;\Delta_{\phi}(A+B)\right]^{2} \nonumber \\
 && + \;\,
 2\;\left(\Delta_{\phi}A\right)^{2}\;\cdot\;\left(\Delta_{\phi}B \right)^{2}. \label{n13b}
\end{eqnarray}

The last three inequalities represent  new upper bounds on the sum of variances 
and they  look like a new form of the reverse uncertainty relations for such a  sum of variances or squares of variances.

\section{Limiting cases}

Now let's check what happens with the inequality (\ref{Pat4}) when $|\phi\rangle$ is an eigenstate, e.g. of the operator $B$, i.e. when $|\phi\rangle = |\psi_{b}\rangle$.
 In such a case, relations (\ref{B=b}) and (\ref{B=b+1}) are satisfied,  which means that property  (\ref{DBb=0}) takes place,  and by virtue of (\ref{d(A+B)}) we obtain
\begin{equation}
\delta_{\psi_{b}}( A \pm B) | \psi_{b} \rangle  \equiv \delta_{\psi_{b}} A | \psi_{b} \rangle  \pm \delta_{\psi_{b}} B| \psi_{b}\rangle \equiv
 \delta_{\psi_{b}} A| \psi_{b}\rangle. \label{dA+dB=dA}
\end{equation}
The elementary conclusion following from this last relation is that the right hand--side of the inequality (\ref{Pat4}) calculated for $|\phi\rangle = |\psi_{b}\rangle$ looks as follows
\begin{equation}
\Delta_{\psi_{b}}(A + B) \equiv  \Delta_{\psi_{b}} A, \label{DA+DB=DA}
\end{equation}
and consequently the inequality (\ref{Pat4}) takes the form,
\begin{equation}
 \Delta_{\psi_{b}} A \geq   \Delta_{\psi_{b}} A. \label{Pat4=0}
\end{equation}
Thus, if the quantum system under study is in a state that is an eigenstate of one of the observables, e.g. $B$, then, similarly to the HR uncertainty relation, the sum uncertainty relation (\ref{Pat4}) does not provide any information about a lower bound on the standard deviation $ \Delta_{\psi_{b}} A$ of the observable $A$.


We will now focus on the analysis of the properties of the inequality (\ref{M12}),
or more precisely, on the study of the equivalent inequality (\ref{m2}) preceding it.
Let us now repeat our calculations performed for the Heisenberg-- Robertson uncertainty principle and for the sum uncertainty relation (\ref{Pat4})
in the case where $|\phi\rangle$ is an eigenvector of the operator, e. g. $B$, and it is not an eigenvector of $A$.
Again, using results
(\ref{B=b}), (\ref{B=b+1}),  (\ref{DBb=0}) and (\ref{d(A+B)}) we can conclude that the result (\ref{DA+DB=DA}) takes also place in the considered case.
This in turn means that in this case the inequality (\ref{M12})
  takes the following form (see also, e.g., \cite{ku-edu})
\begin{equation}
	(\Delta_{\psi_{b}} A)^{2}  \geq	 \frac{1}{2} \left( \Delta_{\psi_{b}} A \right)^{2}. \label{ku1}
\end{equation}
Remember, by definition (\ref{DF}) there is $\Delta_{\psi_{b}}(A) \geq 0$. If  $\Delta_{\psi_{b}}(A) = 0$ the inequality  (\ref{ku1}) takes the form $0=0$. On the other hand, if  $\Delta_{\psi_{b}}(A) > 0$
then we have  two absolutely trivial solutions of the this inequality:
The first one: $ 1 > 1/2$, and the second one:
\begin{equation}
	(\Delta_{\psi_{b}}A)^{2}  \geq	0, \label{DAb>0}
\end{equation}
which is already known from the definitions (\ref{DF}) and (\ref{dF}).
Again, the last result is the same as the one that, which follows from the Heisenberg--Robertson uncertainty principle.

We see that, in the case when the system is in state $|\phi\rangle$ which is an eigenstate of, for example, operator $B$, $|\phi\rangle = |\psi_{b}\rangle$, in contrast to the HR uncertainty relation (\ref{R1}), (see (\ref{[A,B]=0})),  the right-hand sides of the sum uncertainty relations (\ref{Pat4=0}) and (\ref{ku1}) are non--zero. However, this does not mean that these relations are non-trivial. Simply, in this case these sum uncertainty relations are trivially fulfilled for any $\Delta_{\psi_{b}}A \geq 0$.


Now let's check what happens with inequalities  (\ref{rev1a}) and (\ref{n4})  when the state $|\phi\rangle$ is an eigenstate of operator $B$, i.e. when  $|\phi\rangle = |\psi_{b}\rangle$. Proceeding in this case in the same way as in the case of (\ref{Pat4}) and (\ref{M12}) we get
\begin{equation}
(\Delta_{\psi_{b}}A)^{2} \leq (\Delta_{\psi_{b}}A)^{2},
\end{equation}
for inequality (\ref{rev1a}), and
\begin{equation}
(\Delta_{\psi_{b}}A) \geq \left|\Delta_{\psi_{b}}A\right|
\end{equation}
for inequality (\ref{n4}). (Here  results (\ref{dA+dB=dA}), (\ref{DA+DB=DA}) and (\ref{DBb=0}) were used). Similar results can be obtained for the inequality (\ref{Bar2}). The above procedure applied to the inequality (\ref{Bar2}) leads to the following result,
\begin{equation}
(\Delta_{\psi_{b}}A)^{2} \leq (\Delta_{\psi_{b}}A)^{2}. \label{Bar3}
\end{equation}
As can be seen from this, in the case when the state $|\phi\rangle$ is an eigenstate of one of the two observables, $A$ and $B$,  e.g. $B$, from inequalities (\ref{rev1a}), (\ref{n4}) and (\ref{Bar2}) we also do not obtain any significant bounds on the standard deviation, e. g. $\Delta_{\psi_{b}}A$, of the observable $A$. Similar results take place  when $|\phi\rangle$  is an eigenvector of $A$.


It turns out that in the state space one can find vectors that are not eigenvectors of any pair of non--commuting observables $ A$ and $B$, and yet the uncertainty relations analyzed above
do not provide any nontrivial lower or upper bounds on the sums of standard deviations and variances calculated for these vectors, (see, e.g. \cite{ku-min,ku1}).
This property is true when vectors $|\psi_{A}\rangle = \delta_{\phi}A|\phi\rangle$ and $|\psi_{B}\rangle = \delta_{\phi}B|\phi\rangle $  are non--zero and orthogonal to each other: $\langle \psi_{A}|\psi_{B}\rangle \equiv
\langle \phi|\delta_{\phi}A\;\delta_{\phi}B|\phi\rangle =0$ and $|\phi\rangle$ is not an eigenvector of either $A$ or $B$.
Then $\Delta_{\phi} A > 0$ and $\Delta_{\phi}B > 0$.
From an analysis of properties
of  the correlation function ${\cal C}_{\phi}(A,B)$ defined by the formula (\ref{C1}) we conclude that  ${\cal C}_{\phi}(A,B) = 0$  means that
there are no any correlations between observables $A$ and $B$ in the state $|\phi\rangle$ (i.e. they are uncorrelated in this state 
--- here,  and in the subsequent parts  of the paper we employ an interpretation of ${\cal C}_{\phi}(A,B) = 0$ analogous to that in \cite{Mac2}).

So, if ${\cal C}_{\phi}(A,B) =0$ then
$ |\psi_{A}\rangle \perp |\psi_{B}\rangle$
 and in such a case  we have
 \begin{equation}
 \left\| \psi_{A} \rangle \pm |\psi_{B}\rangle \right\|^{2} \equiv \left\| \,|\psi_{A} \rangle \right\|^{2} + \left\|\,|\psi_{B}\rangle \right\|^{2},\;\; {\rm if}\;\; |\psi_{A}\rangle \perp |\psi_{B}\rangle.
 \label{perp}
\end{equation}
If we now assume that $ |\psi_{A}\rangle \perp |\psi_{B}\rangle$ and square both sides of the triangle inequality (\ref{t1}) and apply the result (\ref{perp}), then instead of the inequality (\ref{t1}) we simply get that
\begin{equation}
2\,\left\| \,|\psi_{A} \rangle \right\| \; \left\|\,|\psi_{B}\rangle \right\| \geq 0. \label{perp2}
\end{equation}
Then, remembering the definitions (\ref{dA}), from the last inequality we conclude that the sum uncertainty relation (\ref{Pat4}) in the considered case simplifies to
\begin{equation}
\Delta_{\phi}A\,\cdot\,\Delta_{\phi}B \,\geq\,0.  \label{t-hr}
\end{equation}
Applying the result (\ref{perp}) to the triangle inequality of the second kind (\ref{t2}) and proceeding as in the previous case, we find that the stronger sum uncertainty relation (\ref{M12}) 
can be written as follows
\begin{equation}
(\Delta_{\phi}A)^{2} + (\Delta_{\phi}B)^{2}\,\geq \,0, \label{perp+t2}
\end{equation}
in the considered case. 
(An insignificant factor of $1/2$ has been omitted here).
The common feature of inequalities (\ref{t-hr}), (\ref{perp+t2}) is that they do not provide any significant information about the lower bound on the sum of standard deviations or on the sum of variations. Simply, the information resulting from these results is the same as that known from the definition (\ref{DF}) of the standard deviation $\Delta_{\phi}F$ of the observable $F$.
Similar conclusions hold for observables $A$ and $ B$ uncorrelated in state $|\phi\rangle$ in the case of the remaining uncertainty relations discussed in the previous section.

Note that in the case where,  observables $A$ and $B$ are uncorrelated in state $|\phi\rangle$, i.e. when $|\psi_{A}\rangle = \delta_{\phi}A|\phi\rangle \;\perp\;|\psi_{B}\rangle = \delta_{\phi}B|\phi\rangle $, then equations (\ref{[dA,dB]}) and  (\ref{C00}) imply that $|\langle \phi|[A,B]|\phi\rangle| = 0$.
This property means that that if ${\cal C}_{\phi}(A,B) = 0$, then the right hand--side of the standard HR uncertainty relation (\ref{R1}) is equal  to zero despite the fact that $\Delta_{\phi}A  > 0$  and $\Delta_{\phi}B >0$.
To perform a more in--depth analysis of this property, let us note that the following relation follows from (\ref{[dA,dB]}),  (\ref{Im[]}) and (\ref{C1}),
\begin{equation}
\langle \phi| [A,B] |\phi \rangle \equiv 2 i\,\Im\,[{\cal C}_{\phi}(A,B)]. \label{[A,B]=ImC}
\end{equation}
So, in fact, as can be seen from (\ref{Im[]}) and (\ref{[A,B]=ImC}), it is enough that $\Im\,[ C_{\phi}(A,B)=0$ so that $|\langle \phi|[A,B]|\phi\rangle| = 0$ in the inequality (\ref{R1}).
It should be noted here that in such cases, i.e. when $\Im\,[ {\cal C}_{\phi}(A,B)]= 0$,
users of
the stronger sum uncertainty relation (\ref{M12}) may interpret this fact
as an advantage of the stronger uncertainty relation over the standard HR uncertainty relation (\ref{R1}).
This is because, in this case, inequality (\ref{M12}) 
looks as follows
(see \cite{ku-edu}),
\begin{equation}
(\Delta_{\phi}A)^{2}   + (\Delta_{\phi}B)^{2}   \geq  2|{\cal C}_{\phi}^{clas}(A,B)|.  \label{M12+new}
\end{equation}
(The proof is given in Appendix A).
Thus, in all those cases when $\Im\,[{\cal C}_{\phi}(A,B)]$  $= 0$, which results in that the right side of the HR inequality (\ref{R1}) is equal to zero, the right side of the stronger uncertainty relation (\ref{M12}) in the form (\ref{M12+new}) can be different from zero.
  So, even in such cases, contrary to the standard HR uncertainty relation,   there can be a lower, non--zero, bound for the sum of variances $(\Delta_{\phi}A)^{2} + (\Delta_{\phi}B)^{2}$.
On the other hand, the inequality (\ref{M12+new})  is simply an upper bound on the modulus of the real part of the quantum version of correlation function ${\cal C}_{\phi}(A,B)$, i.e., on modulus of ${\cal C}_{\phi}^{clas}(A,B)$.

  It should be noted here that the above--mentioned ,,advantage'' of the state dependent ,,stronger sum uncertainty relation'' on the state dependent HR uncertainty relation does not occur in the case of the Schr\"{o}dinger--Robertson (SR) uncertainty relation \cite{M,R,S}, which is more sensitive to effects  of type
  ${\cal C}_{\phi}^{clas}(A,B) = \Re[{\cal C}_{\phi}(A,B)] \neq  0$, and $\Im[{\cal C}_{\phi}(A,B)]=0$
   than the HR relation --- see Appendix B. Consequently, it can be stated on the basis of the above considerations that uncertainty relations such as the state dependent sum uncertainty relations or stronger sum uncertainty relation in comparison to HR and SR uncertainty relations, in those cases in which the right side of the HR and SR relations is equal to zero, have no any advantage over HR and SR relations. 
   So,
the apparent superiority of the state dependent sum (\ref{Pat4}), (\ref{t-n1}) and stronger sum (\ref{M12}), (\ref{M12aa}), (\ref{Bar2})
uncertainty relations over HR in the critical cases is an artefact of comparing against the wrong
benchmark; measured against SR relation, it disappears entirely.

\section{Final remarks}

The Heisenberg--Robertson  uncertainty relation (\ref{R1}) as well as sum uncertainty relations (\ref{Pat4}), (\ref{M12}) and others derived above were be proven using mathematically exact inequalities resulting from the properties of Hilbert spaces.
There are many rigorous inequalities in Hilbert space \cite{Cer} which can be used to find various
relations  of the kind discussed above.
Many of them provide us with information about the lower bound on the product or sum of standard deviations, or variances and these lower bounds are studied in many papers.
At this point it is worth mentioning that the properties of Hilbert spaces provide us with the tools to find not only these lower bounds, but also other relations and bounds for the sums of standard deviations or variances.
For example, by inserting $|\psi_{A}\rangle$  and $|\psi_{B}\rangle$ in the form (\ref{dA}) into the parallelogram law (\ref{par}),
 we find that there must be (see \cite{strong-2}):
\begin{equation}
(\Delta_{\phi}A)^{2} + (\Delta_{\phi}B)^{2}  = \frac{1}{2} \left[\left(\Delta_{\phi}(A + B)\right)^{2} + \left((\Delta_{\phi}(A - B)\right )^{2}\right]. \label{DA+DB-5}
\end{equation}
Another example is the inequality (\ref{Deb}), that is an upper bound on the sum of variances which was obtained   in  \cite{Deb} using the Dunkle-Williams inequality \cite{Cer,Dun}.
This example shows that
the geometry of Hilbert space of states of a quantum system forces the sum of variances to have both the lower (\ref{M12}) and the upper (\ref{Deb}) bounds. In the literature, inequalities of the type of (\ref{Deb}) are called the reverse uncertainty relations (see, e.g. \cite{Deb,Xiao2,Xiao3,Xiao4}).

Using inequalities (\ref{n5}) and (\ref{R1}) one finds that (see also \cite{Mac}),
\begin{equation}
\frac{1}{2}\left( (\Delta_{\phi}A)^{2} + (\Delta_{\phi}B)^{2} \right) \;\geq \;\Delta_{\phi}A\;\Delta_{\phi}B\; \geq\;\frac{1}{2} \left|\langle [A,B] \rangle_{\phi} \right|. \label{n6}
\end{equation}
Similarly, an analogous chain of inequalities can be easily found for the sum of variances. Namely, combining inequalities (\ref{rev1a}) and (\ref{M12}), we find
\begin{equation}
\left( \Delta_{\phi}(A-B) \right)^{2} + 2 \Delta_{\phi} A\; \Delta_{\phi} B\,\geq \,
(\Delta_{\phi}A)^{2} + (\Delta_{\phi}B)^{2} \, \geq  \,  \frac{1}{2}\,  [\Delta_{\phi}(A +B)]^{2}. \label{nv}
\end{equation}
On the other hand, if to use relations (\ref{rev1a}) and (\ref{n5}) one obtains
\begin{equation}
\left( \Delta_{\phi}(A-B) \right)^{2} + 2 \Delta_{\phi} A\; \Delta_{\phi} B\,\geq \,
(\Delta_{\phi}A)^{2} + (\Delta_{\phi}B)^{2} \, \geq  \,   2 \Delta_{\phi}A\;\Delta_{\phi}B. \label{nv1}
\end{equation}

Equation (\ref{DA+DB-5}) implies that, 
the left--hand side of the  revers  uncertainty relation (\ref{Deb}) can be replaced by the right--hand side of equation (\ref{DA+DB-5}).
This, in turn,  means that the right--hand side of inequality (\ref{Deb}) also serves as an upper bound for the expression appearing on the right--hand side of equation (\ref{DA+DB-5}). In other words, upper bounds for a sum of varaiances and for $ \frac{1}{2} [\left(\Delta_{\phi}(A + B)\right)^{2} + \left((\Delta_{\phi}(A - B)
\right )^{2}]$ must be the same. The same can be said about the lower bound for the sum of variances. The lower bound for  $\left( (\Delta_{\phi}A)^{2} + (\Delta_{\phi}B)^{2} \right) $ 
is also a lower bound for $ \frac{1}{2} [\left(\Delta_{\phi}(A + B)\right)^{2}$ \linebreak $ + \left((\Delta_{\phi}(A - B)
\right )^{2}]$, and vice versa.
Strictly speaking: Equation  (\ref{DA+DB-5}) states that any inequality satisfied by the expression on its left--hand side is also satisfied by the expression on its right--hand side, and vice versa.

For the reader's information: 
The implications of the uncertainty equality (\ref{DA+DB-5}) for the properties of the Maccone--Pati stronger sum uncertainty relation were studied in  \cite{strong-2}.
Therein two other uncertainty equalities were derived, and the uncertainty relations resulting from them, including the stronger sum uncertainty relation, were studied and analyzed.

The uncertainty relations and other inequalities analyzed above depend on the state of the quantum system under study --- more precisely, on a vector from the Hilbert space of the system's states.
They describe the theoretical dispersion of values that measured observables can teke when their values are determined across a very large number of identical systems, all in the same given state.
These measurements are ideal quantum measurements, and by assumption, they  do not disturb the system being studied.
Such relations do not account for the specific characteristics or the precision of the real measuring instruments used.
Measurements performed with real measuring instruments will always result in a greater spread of results than that predicted by the aforementioned uncertainty relations; the spread of results arising from these uncertainty relations serves merely as a lower bound for the spread caused by the use of real measuring instruments.


As mentioned, using some  inequalities valid in the Hilbert state space, one can find various relations between standard deviations or variances of non--commuting observables.
These can be
lower bounds on products of variances or standard deviations, or lower or upper bounds on their sums as well as upper bounds (\ref{n13}), (\ref{n13a}) and (\ref{n13b}).
There can also be relations similar to Eq. (\ref{DA+DB-5}). It seems, that relations of the type discussed above may be useful, e.g., in quantum metrology. In general, their usefulness, as well as the usefulness of the relations discussed in Sec. 4, depends on the goal we want to achieve, or the phenomena we want to study, describe  and understand. For this reason, all relations of the mentioned types require further research.

\hfill\\
\section*{Appendix A}
Let us analyze properties of the term $[\Delta_{\phi}(A+B)]^{2} $ appearing on the right--hand side of the stronger uncertainty relation (\ref{M12}).
 Using (\ref{dF}) and (\ref{d(A+B)}),
 after some algebra,  we have,
 $$
 [\Delta_{\phi}(A+B)]^{2} = \|\, \delta_{\phi} A|\phi\rangle + \delta_{\phi}B|\phi\rangle \|^{2} 
  \equiv [\Delta_{\phi}A]^{2} + [\Delta_{\phi}B]^{2} + 2 \Re\,[{\cal C}_{\phi}(A,B)] \geq 0. \eqno(A1)
 $$
This means that
$$
 [\Delta_{\phi}A]^{2} + [\Delta_{\phi}B]^{2}\, \geq \, -\,2 \Re\,[{\cal C}_{\phi}(A,B)]. \eqno(A2)
 $$
On the other hand, using $(A1)$ we can rewrite the stronger uncertainty relation (\ref{M12}) as follows
$$
 [\Delta_{\phi}A]^{2} + [\Delta_{\phi}B]^{2}\, \geq \,\frac{1}{2} [ \Delta_{\phi}(A+B)]^{2} \equiv \frac{1}{2} \{ [\Delta_{\phi}A]^{2} + [\Delta_{\phi}B]^{2} +  2 \Re\,[{\cal C}_{\phi}(A,B)]  \},
$$
which can be rewritten as
$$
[\Delta_{\phi}A]^{2} + [\Delta_{\phi}B]^{2}\, \geq \, +\,2 \Re\,[{\cal C}_{\phi}(A,B)]. \eqno(A3)
$$
Now, if $\Re\,[{\cal C}_{\phi}(A,B)]$ is negative: i.e., if $ \Re\,[{\cal C}_{\phi}(A,B)] = - |\Re\,[{\cal C}_{\phi}(A,B)]|$, then from $(A2)$ we get
$$
[\Delta_{\phi}A]^{2} + [\Delta_{\phi}B]^{2}\, \geq \, +\,2 |\Re\,[{\cal C}_{\phi}(A,B)]|. \eqno(A4)
$$
The inequality ($A3$) also is correct in this case because in any case the left hand side of ($A3$) is non--negative.

If  $\Re\,[{\cal C}_{\phi}(A,B)]$ is positive: i.e., if $ \Re\,[{\cal C}_{\phi}(A,B)] = + |\Re\,[{\cal C}_{\phi}(A,B)]|$, then it is obvious that Eq. ($A2)$ holds and that  the Eq. ($A3$) leads to the conclusion that again
$$
[\Delta_{\phi}A]^{2} + [\Delta_{\phi}B]^{2}\, \geq \, +\,2 |\Re\,[{\cal C}_{\phi}(A,B)]|. \eqno(A5)
$$
Results ($A4$) and ($A5$)  show that  the inequality (\ref{M12+new}) hold.

\section*{Appendix B}
Starting from Schwarz's inequality,
$$
\|\,|\psi_{A}\rangle\| \, \|\,|\psi_{B}\rangle \| \geq |\langle \psi_{A}|\psi_{B}\rangle|, \eqno(B1)
$$
and replacing $|\psi_{A}\rangle, |\psi_{B}\rangle$ by vectors defined in Eq. (\ref{dA}), and then using definitions (\ref{DF}), (\ref{dF})   and (\ref{C1})
we get
$$
\Delta_{\phi}A\; \Delta_{\phi}B\;\geq\; |{\cal C}_{\phi}(A,B)|. \eqno(B2)
$$
This relation can be rewritten as follows
$$
[\Delta_{\phi}A]^{2}\; [\Delta_{\phi}B]^{2}\;\geq\; \{\Re[{\cal C}_{\phi}(A,B)]\}^{2} + \{\Im[{\cal C}_{\phi}(A,B)]\}^{2}, \eqno(B3)
$$
what is the compact form of the SR uncertainty  relation.
From this relation and from (\ref{[A,B]=ImC}) it can be seen that the lower bound on the  product of variances $[\Delta_{\phi}A]^{2}$ and  $[\Delta_{\phi}B]^{2}$ does not have to be equal to zero when $\Im[{\cal C}_{\phi}(A,B)]=0$, i. e. when the lower bound resulting from the HR uncertainty relation is zero. Furthermore, as the derivation shows, the  inequality $(B3)$ holds for both non--commuting and commuting operators. In the case where $[A, B] = 0$, inequality $(B3)$ can be interpreted as an upper bound on modulus of the classical part ${\cal C}_{\phi}^{clas}(A,B) $  of the correlation function ${\cal C}_{\phi}(A,B) $:
$$
\Delta_{\phi}A\; \Delta_{\phi}B\;\geq\; \left|
\Re[{\cal C}_{\phi}(A,B)]\right| \equiv  \left| {\cal C}_{\phi}^{clas}(A,B) \right|. \eqno(B4)
$$

\noindent
{\bf Acknowledgements:} The author wishes to express sincere thanks to the anonymous reviewer for all the comments and suggestions that helped improve and refine the work.

\section*{Declarations}

\begin{itemize}
\item This work was supported by
the program of the Polish Ministry
of Science and Higher Education under the name "Regional
Initiative of Excellence", Project No. RID/SP/0050/2024/1.

\item  The author declares no conflict of interest.

\item The author declares that there is not any personal, academic interest, or any
other factors that may be perceived to influence the objectivity, integrity or
value of the study.

\item The author consents to publication. 

\item This manuscript has no associated data, or the data will not be deposited.
[Author's comment: This is a theoretical work and analytical calculations
are made. Therefore, no data are required].
\end{itemize}

\end{document}